# The Role of Lifetime Exposures across Cognitive Domains in Barbados Using Data From the SABE Study

Jason Steffener[1*], Joanne Nicholls[2], Dylan Franklin[3]

1 Interdisciplinary School of Health Science, University of Ottawa, Ottawa, ON Canada
2 Teachers College, Columbia University, New York City, NY, USA
3 School of Psychology, University of Ottawa, Ottawa, ON, Canada

* Corresponding Author

Jason Steffener, PhD
Associate Professor
Interdisciplinary School of Health Sciences
University of Ottawa,
200 Lees, Lees Campus,
Office # E250E,
Ottawa, Ontario, CANADA
K1S 5S9
Email: jsteffen@uottawa.ca

# Abstract

This study characterized the effects of aging on individual cognitive domains and how sex, job type, and years of education alter the age effect on older adults from Barbados. This was an analysis of the cross-sectional data collected as part of the SABE Study (Health, Well-being and Ageing) in 2006. The loss of a single point in each of the individual cognitive domains assessed using the mini-mental state exam served as dependent variables. Independent variables included age, sex, years of education, job type, and the interactions with age in a series of logistic regression analyses. The study aimed to identify which factors altered the effect of age on cognitive performance and which directly affected performance. Results demonstrated that the effect of age differed across the cognitive domains. In addition, sex, education, and job type all differentially affected cognitive performance in an additive, formative manner. The most consistent finding was that high years of education coupled with employment requiring mostly mental effort was the best combination for maintaining high levels of cognitive performance in late life. The results demonstrate that adverse age effects on cognitive performance may be minimized or delayed through modifiable lifetime exposures in the people of Barbados.

# Introduction

Advancing age has a large effect on cognitive decline, and it is also one of the greatest risk factors for future development of Alzheimer's disease (Riedel et al., 2016; Salthouse, 2019). Fortunately, certain lifetime exposures and well-being benefit cognitive performance and offer some protection later in life (Wang et al., 2019). Protection in its simplest form delays the onset of disease, thereby avoiding a disease entirely. Neurological protections include the maintenance of brain tissue, the build-up of neural reserves, or the ability to adapt neural functions in response to pathological neural processes. Beneficial lifetime exposures include education, occupational complexity, engagement in leisure activities, greater social networks, socioeconomic status, and bilingualism (Bak et al., 2014; Bielak et al., 2019; Farina et al., 2018; Harrison et al., 2015; Klimova & Dostalova, 2020).

Studies of the neural protection of lifetime exposures fall into two broad categories: brain reserve and cognitive reserve (Cabeza et al., 2018; Steffener et al., 2014; Stern et al., 2020). The theory of brain reserve describes the accumulation of neural resources due to genetic, environmental, or lifestyle factors to mitigate the negative effects of aging. The theory of cognitive reserve states that the accumulation of functional resources due to genetic, environmental, or lifestyle factors allows brain function to adapt to the negative effects of aging through the use of functional compensatory resources. In the absence of neural measures, studies of the beneficial effects of lifetime exposures on cognition use age as a predictor.

Beneficial lifetime exposures provide neural and functional protection from aging effects. In addition, they are also directly related to cognitive performance (Correa Ribeiro et al., 2013; Steffener et al., 2014) and limit the negative effects of age on cognitive decline. In this way, beneficial exposures directly influence cognitive performance. Their effect on limiting the influence of age on cognitive performance may be of greater influence on delaying clinically significant cognitive decline (Dekhtyar et al., 2015; González et al., 2013; Wang et al., 2019).

Educational attainment is one of the most explored lifetime exposures related to cognitive performance (Opdebeeck et al., 2016; Stern, 2002, 2009). Greater years of education demonstrate a wide array of benefits, such as higher levels of cognitive performance than those with fewer years of education. There are also indirect effects that higher education offers someone, which include better occupations, better access to healthcare, higher vocabulary skills, et cetera. In the SABE large-scale study of older adults in the Americas (Pelaez et al., 2006), education demonstrated a beneficial effect on general cognitive performance across multiple countries, including Argentina, Mexico, Uruguay, Chile, and Brazil (Maurer, 2011). However, there was no evidence of beneficial educational effects when using general cognition as the output in the two countries with the highest literacy rates and educational attainment, Barbados and Cuba. The current work focuses on the English-speaking country of Barbados to investigate the role of lifetime exposures on cognitive subdomains instead of their amalgamation.

Tests of general cognition often assess a combination of multiple cognitive domains, intending to screen for a cognitive decline of any form in any domain. A limitation of general cognitive screening assessments is their design to be sensitive and quick to administer. This focus on total scores limits any insight into differential aging effects on the various cognitive domains (Scuteri et al., 2005). As we age, cognitive processing speed typically declines first. It has the most rapid decline, while crystallized intelligence, often measured as vocabulary skills, increases across the lifespan and does not decline until much later (Salthouse, 2019). Composite total scores, therefore, miss changes in cognitive subdomains. The current project characterized the effects of aging on various cognitive domains and how sex, job type, and years of education alter the age effect on older adults from Barbados.

## Methods

### Data

The cross-sectional data for this study are from the Survey on Health, Well-Being and Ageing (SABE) in Latin America and the Caribbean (Pelaez et al., 2006). The SABE is a multi-city study conducted in seven Latin American and Caribbean cities, including Bridgetown, Barbados. The sample collected in Barbados was collected through a multi-stage stratified cluster sampling approach. First, using the national election registry, a random sample of households was selected. One individual aged 60 years and older was selected and interviewed within each household. Of the 1,878 households identified with a person above sixty, there was an 80% response rate, with 1,508 interviews conducted. The interview assessed cognitive abilities, sociodemographics, household characteristics, health and functional status, health care utilization, and anthropometry.

The present study sample included 1325 older adults with no measured cognitive deterioration and complete data. Complete data means each participant had values for age in years, job type, sex, education in years, and scores on the Mini-Mental State Exam (MMSE) (Folstein et al., 1975). Missing data included: 44 for MMSE, 13 for education, and 76 for the job type. As in the original study, cognitive deterioration is an MMSE score of 12 or below and accounted for 55 participants.

### Dependent Variables

The nine-variable, 19-point version of the mini-mental state exam (MMSE), was used in this study (Folstein et al., 1975). A World Health Organization developed this shortened version using regression analysis to identify the questions that best explain cognitive deterioration (Icaza & Albala, 1999). The MMSE consists of six domains: **orientation** assessed with questions about the time (four points), **immediate memory recall** of three words (three points), **working memory** assessed with a reverse digit span of five digits (five points), following three **instructions** when handed a piece of paper (three points), **delayed memory recall** of the three words from earlier in the exam (three points), and **visuospatial abilities** assessed with the drawing of two intersecting circles (one point). Each cognitive domain predicted the loss of one point; therefore, each model only included participants who had all points or the loss of one point in each domain.

**Independent Variables**

The independent variables consisted of age, years of education, job type, and sex. The mean age was 71.82 years, with a standard deviation of 7.81 years and a range of 60 to 97 years. The mean years of education were 5.39 years, with a standard deviation of 3.44 years and a range of 0 to 20. There were multiple modes in the data, with 25 people having zero years of education, 112 having one year, 318 having three years, 529 having five years, 170 having ten years, and 29 having 15 years. The range was zero to twenty years of education. Assessment of job type used self-description of jobs mainly requiring physical effort (N = 897), requiring mental effort (N = 216), or a mixture of the two (N = 212). Work largely involving physical effort included laborers, agricultural workers, and service workers. Work largely involving mental effort included professionals, executives, and office workers. Mixed work included sales clerks, nurses, and laboratory technicians. The sex of participants included 786 females and 539 males.

**Data Analysis**

Binomial logistic regression was performed for each cognitive domain using the Jamovi v1.6 software package (Jonathon Love, 2019). A priori tests of the linearity of the continuous predictor variables, education, and age, were tested using the Box-Tidwell test (Box & Tidwell, 1962). In addition, collinearity statistics were assessed for any collinearity between the predictors in the models. The continuous variables were mean-centered to remove collinearity arising when forming interaction terms.

For each cognitive domain, a series of three nested models were compared. The first step included age alone; the second added the covariates of education, job type, and sex. The third step added the interaction of age and each of the three covariates. The most complicated model was selected if it was significant using the overall Chi-squared model test, had the smallest AIC, and represented a significantly better model than the less complicated model using the Chi-squared test for model comparison. If the most complicated model was not selected, the model with only the main effects compared to the model including only the age effect using the same criteria was used.

Once a model was selected, the variables were assessed for significance using omnibus likelihood tests. Model coefficients were assessed for significance using the Wald statistic, and probabilities of cognitive decline were assessed for the interpretation of results. Female sex and physical effort job type were the reference categories in each regression model.

## Results

Table 1 shows the number of participants with scores at the maximum for each cognitive domain and the number that had lost one point for that domain.

Table 1. Sample sizes per domain for those with maximum scores and having had lost one point

|  | N at Max Score | N who Lost One Point | Max Value |
|---|---|---|---|
| Orient | 1172 | 133 | 4 |
| Immediate Recall | 1312 | 12 | 3 |
| Digit Span | 976 | 154 | 5 |
| Paper | 1249 | 66 | 3 |
| Delayed Recall | 844 | 312 | 3 |
| Circle | 1177 | 148 | 1 |
| MMSE Total | 569 | 287 | 19 |

Table 2 shows the results from the six binomial logistic regression models. A brief description of the overall results is provided here, followed by detailed results for each cognitive domain. The model predicting paper was the only domain where the model, including the interaction terms, was significant. The models predicting orientation, digit span, delayed recall, and circle drawing were significant, including age and lifetime exposure variables. No model was significant when predicting immediate recall.

Table 2. Logistic regression parameter estimates for the three nested models and each cognitive domain

| | Intercept | Age | Edu | Job1 | Job2 | Sex | Age:Edu | Age:Job1 | Age:Job2 | Age:Sex |
|---|---|---|---|---|---|---|---|---|---|---|
| Orient | -2.33*** | 0.041*** | -0.15*** | -0.27 | -0.027 | 0.078 | | | | |
| Immediate Recall | -- | -- | | | | | | | | |
| Digit Span | -1.95*** | 0.029** | -0.099** | -0.73* | -0.78** | 0.38* | | | | |
| Paper | -2.94*** | 0.066* | 0.13*** | -1.20* | 0.083 | 0.082 | 0.0081 | -0.0046 | 0.014 | -0.097** |
| Delayed Recall | -0.84*** | -0.0077 | -0.061** | -0.44* | -0.23 | -0.098 | | | | |
| Circle | -2.31*** | 0.086*** | -0.13*** | -0.83* | -0.60* | 0.20 | | | | |

Notes: A negative parameter indicates a decreased probability of losing one point. Job1 refers to mental effort vs. physical effort. Job2 refers to mixed effort versus physical effort. Sex refers to male vs. female. †p<.10; *p<.05; **p<.01; ***p<.001

The size of the age effect varied between each cognitive domain; see Figure 1. This figure plots the probability of losing one point in the next year for each cognitive domain. Except for immediate and delayed recall, all domains demonstrate significant positive age effects. Therefore, for every year older someone is, the probability that they will lose one point increases.

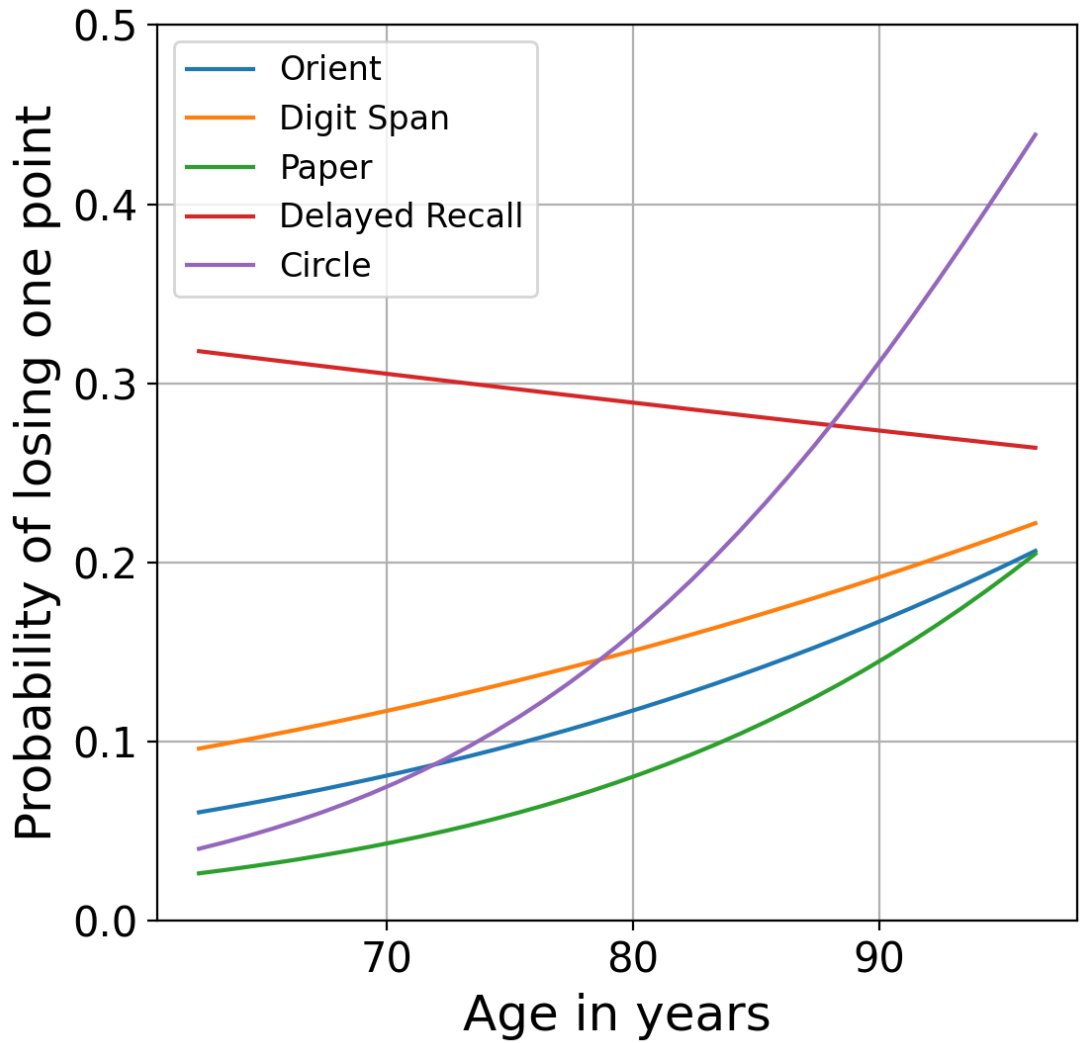

Figure 1. Line plot showing the age effects on the probability of losing one point for each domain. The immediate recall task did not show a significant age effect and is not shown.

Education had significant negative effects on the orient, digit span, delayed recall, and circle domains. Therefore, those participants with more years of education had a lower probability of losing one point. The paper domain demonstrates a significant positive effect of education where more years of education was related to a greater probability of losing one point. The effect of job type was significant for digit span, paper, delayed recall, and circle. In all four domains, greater mental effort within one's job is related to significantly better cognitive scores than requiring mostly physical effort. For the digit span and circle tasks, the mixed job type was significantly better than the physical effort types. Sex significantly modulated the age effect of the paper task. Sex also demonstrated a significant main effect on the digit span working memory domain, where females had a higher probability of losing one point. These effects are plotted for each domain in Figures 2 to 6. There is no plot for immediate recall because of the lack of significant relationships.

**Orientation**

The model containing interaction terms was not significantly better than the model containing the main effects of the covariates ($X^2(4) = 0.412$, $p = 0.981$). The model with the main effects was significantly better than the model containing only age ($X^2(4) = 24.0$, $p < 0.001$) and was significant using the overall model test ($X^2(5) = 41.4$, $p < 0.001$). This model also had the lowest AIC of the three models.

The omnibus likelihood ratio tests showed that age ($X^2(1) = 12.6$, $p < 0.001$) and education ($X^2(1) = 16.8$, $p < 0.001$) were significant; however, sex ($X^2(1) = 0.167$, $p = 0.682$) and job type ($X^2(2) = 0.610$, $p = 0.737$) were not. The unstandardized estimates for the intercept (B = -2.33, SE = 0.173, Wald = 182, $p < 0.001$), age (B = 0.0410, SE = 0.0115, Wald = 12.7, $p < 0.001$) and education (B = -0.151, SE = 0.0393, Wald = 14.9, $p < 0.001$) were significant. The unstandardized estimates for sex (B = 0.0782, SE = 0.192, Wald = 0.167, $p = 0.683$), and both job type variables mental - physical (B = -0.266, SE = 0.350, Wald = 0.581, $p = 0.446$) and mixed - physical (B = -0.0274, SE = 0.259, Wald = 0.0111, $p = 0.916$) were non-significant. Figure 2 plots these results.

For age, the estimated odds ratio favored an increase of 4.2% (95% CI: 1.02, 1.06) probability of losing one point on the orientation questions for every year older someone was. For education, the estimated odds ratio favored a decrease of 14.1% (95% CI: 0.763, 0.928) probability of losing one point on the orientation questions for every additional year of education someone had.

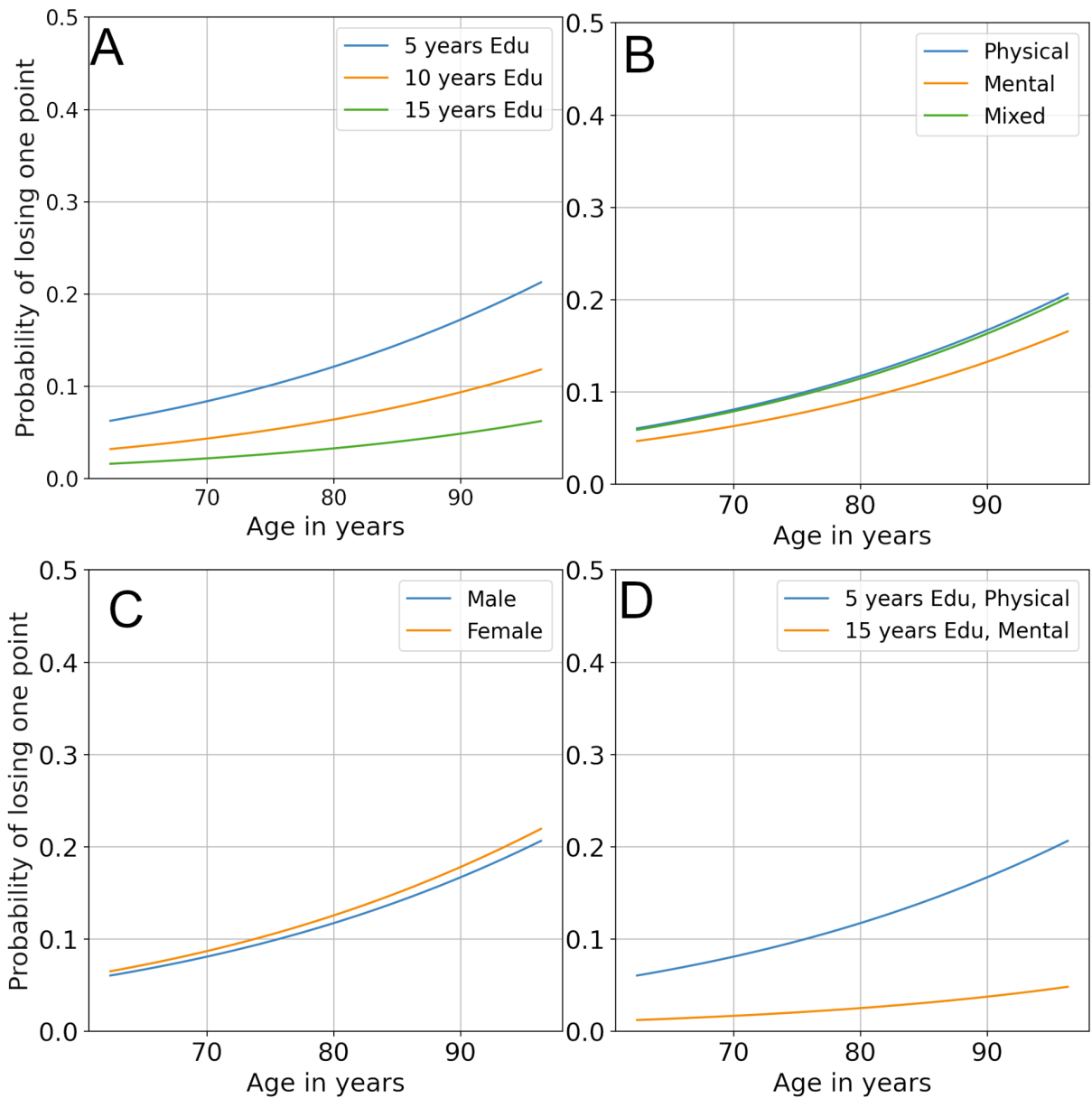

Figure 2. Probability of losing one point on the orientation subscale. Panel A shows the variability across years of education, B) across job types, C) across sex, and D) the combination of the significant effects of education and job type.

**Immediate Recall**

None of the models predicting immediate recall were significant.

**Digit Span**

The model containing interaction terms was not significantly better than the model containing the main effects of the covariates ($X^2(4) = 0.492$, $p = 0.974$). The model with the main effects was significantly better than the model containing only age ($X^2(4) = 36.5$, $p < 0.001$) and was significant using the overall model test ($X^2(5) = 46.2$, $p < 0.001$). This model also had the lowest AIC of the three models.

The omnibus likelihood ratio tests showed that age ($X^2(1) = 6.74$, $p = 0.009$), education ($X^2(1) = 8.60$, $p = 0.003$), job type ($X^2(2) = 11.74$, $p = 0.003$), and sex ($X^2(1) = 4.29$, $p = 0.038$) were significant. The unstandardized estimates for the intercept (B = -1.95, SE = 0.164, Wald = 141, $p < 0.001$), age (B = 0.0293, SE = 0.0112, Wald = 6.82, $p = 0.009$), education (B = -0.0994, SE = 0.0352, Wald = 7.96, $p = 0.005$), both job type variables mental - physical (B = -0.726, SE = 0.336, Wald = 4.66, $p = 0.031$) and mixed - physical (B = -0.782, SE = 0.291, Wald = 7.21, $p = 0.007$), and sex (B = 0.380, SE = 0.186, Wald = 4.18, $p = 0.041$) were all significantly large. Figure 3 plots these results.

For age, the estimated odds ratio favored an increase of 3.0% (95% CI: 1.01, 1.05) probability of losing one point on the digit span questions for every year older someone was. For education, the estimated odds ratio favored a decrease of 9.5% (95% CI: 0.845, 0.970) probability of losing one point on the digit span questions for every year of education someone had. For job type, the estimated odds ratio favored a decrease of 52.7% (95% CI: 0.250, 0.935) probability of losing one point on the digit span questions when having a mental effort-based job as compared to a physical effort type job. For a mixed-type job, the estimated odds ratio favored a decrease of 56.3% (95% CI: 0.258, 0.809). For sex, the estimated odds ratio favored an increase of 46.2% (95% CI: 1.02, 2.10) probability of losing one point on the digit span questions for females compared to males.

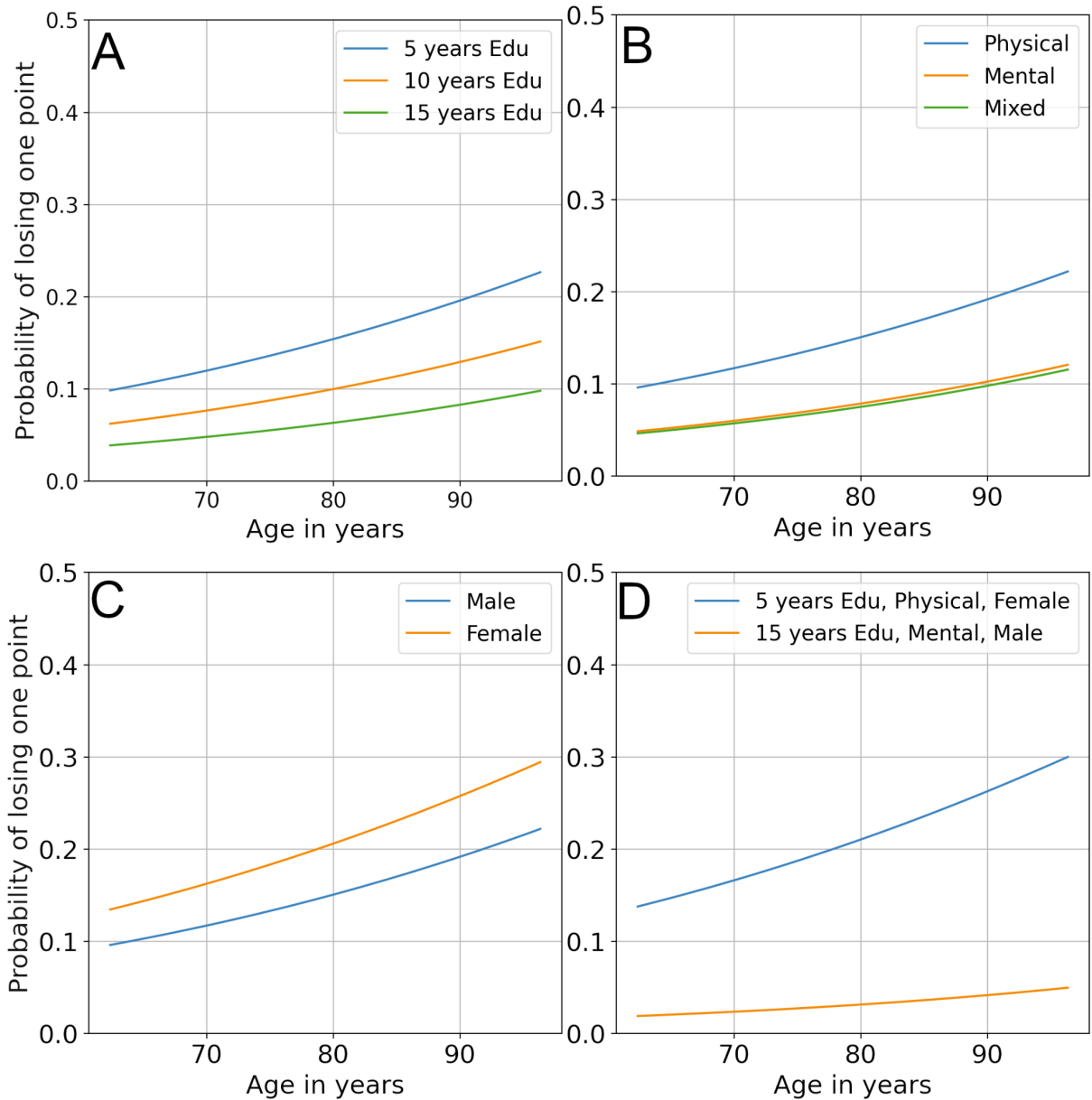

Figure 3. Probability of losing one point on the digit span task. Panel A shows the variability across years of education, B) across job types, C) across sex, and D) the combination of the significant effects of education, job type, and sex.

**Paper**

The model containing interaction terms was significantly better than the model containing the main effects of the covariates ($X^2(4) = 10.70$, p = 0.030) and was significant using the overall model test ($X^2(9) = 24.5$, p = 0.004). This model also had the lowest AIC of the three models. The omnibus likelihood ratio tests showed that the age by sex interaction ($X^2(2) = 7.96$, p = 0.005, age ($X^2(1) = 5.13$, p = 0.024), education ($X^2(1) = 10.99$, p < 0.001), and job type ($X^2(2) = 8.33$, p = 0.016) were significant.

The unstandardized estimates for the intercept (B = -1.95, SE = 0.164, Wald = 141, p < 0.001), age (B = 0.0293, SE = 0.0112, Wald = 6.82, p = 0.009), education (B = -0.0994, SE = 0.0352, Wald = 7.96, p = 0.005), both job type variables mental - physical (B = -0.726, SE = 0.336, Wald = 4.66, p = 0.031) and mixed - physical (B = -0.782, SE = 0.291, Wald = 7.21, p = 0.007), and sex (B = 0.380, SE = 0.186, Wald = 4.18, p = 0.041) were all significantly large. For age, the estimated odds ratio favored an increase of 3.0% (95% CI: 1.01, 1.05) probability of losing one point on the digit span questions for every year older someone was.

For education, the estimated odds ratio favored a decrease of 9.5% (95% CI: 0.845, 0.970) probability of losing one point on the digit span questions for every year of education someone had. For job type, the estimated odds ratio favored a decrease of 52.7% (95% CI: 0.250, 0.935) probability of losing one point on the digit span questions when having a job requiring mental effort as compared to physical effort. For a mixed-effort job, the estimated odds ratio favored a decrease of 56.3% (95% CI: 0.258, 0.809). For sex, the estimated odds ratio favored an increase of 46.2% (95% CI: 1.02, 2.10) probability of losing one point on the digit span questions for females compared to males.

The results from this task, as shown in Figure 4, the results from this task demonstrate age effects in opposite directions for sex. It must be stressed that this plot is of marginal means for only part of the model where other effects are also significant. The final panel of the figure demonstrates that the overall effects between the sexes are similar and differ only in magnitude.

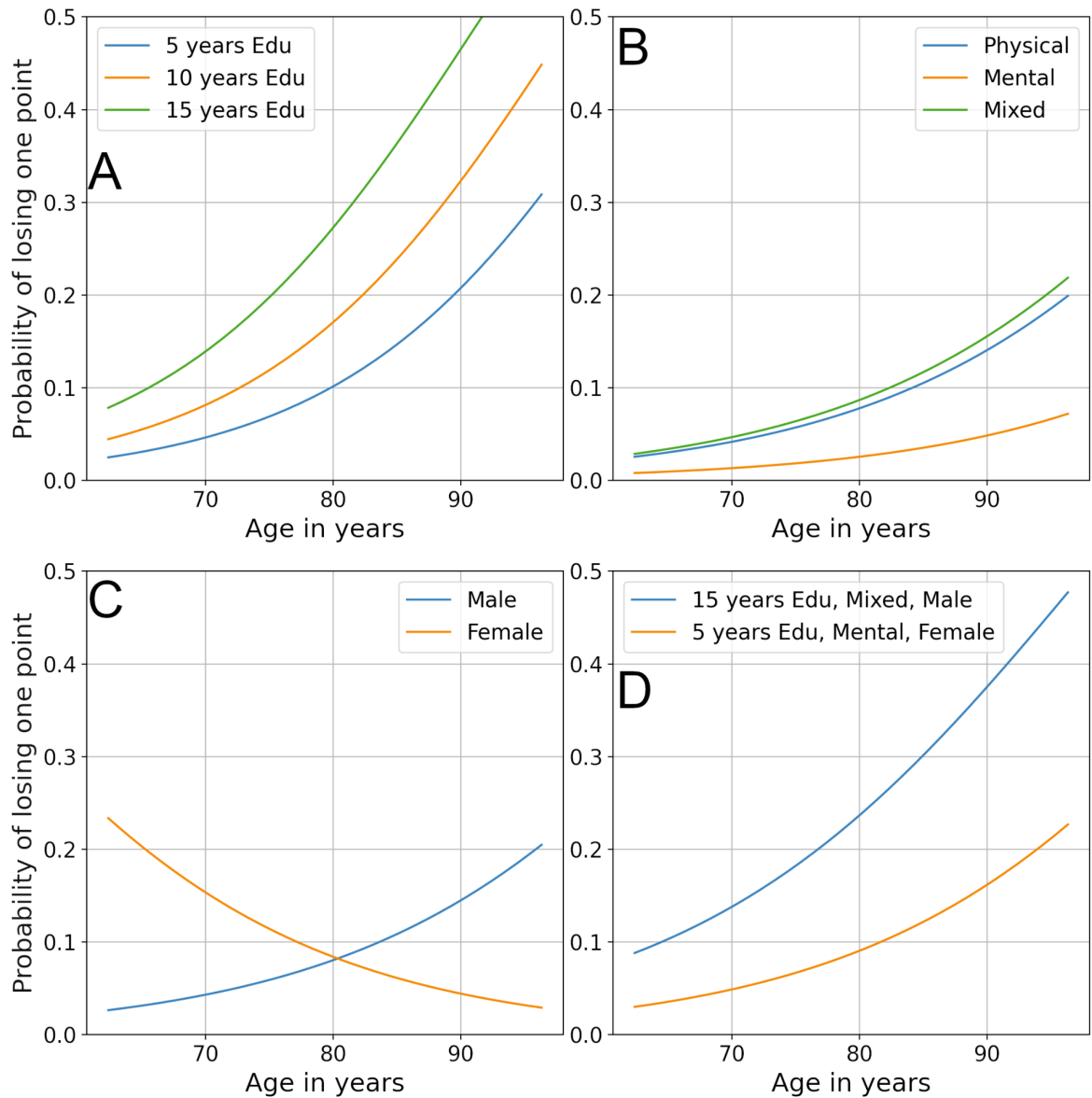

Figure 4. Probability of losing one point on the paper task involving following directions. Panel A shows the variability across years of education, B) across job types, C) across sex, and D) the combination of the significant effects of education, job type, and sex.

**Delayed Recall**

The model containing interaction terms was not significantly better than the model containing the main effects of the covariates ($X^2(4) = 3.64$, p = 0.458). The model with the main effects was significantly better than the model containing only age ($X^2(4) = 21.8$, p < 0.001) and was significant using the overall model test ($X^2(5) = 21.8$, p < 0.001). This model also had the lowest AIC of the three models.

The omnibus likelihood ratio tests showed that education ($X^2(1) = 7.67$, p = 0.0056) was significant; however, age ($X^2(1) = 0.641$, p = 0.423), sex ($X^2(1) = 0.364$, p = 0.546) and job type ($X^2(2) = 4.49$, p = 0.106) were not. The unstandardized estimates for the intercept (B = -0.848, SE = 0.116, Wald = 53.8, p < 0.001), education (B = -0.0636, SE = 0.0234, Wald = 7.39, p = 0.0066), and job type of mental - physical effort (B = -0.431, SE = 0.218, Wald = 3.90, p = 0.0482) were significant. The unstandardized estimates for age (B = -0.0069, SE = 0.0087, Wald = 0.638, p = 0.424), sex (B = -0.0821, SE = 0.136, Wald = 0.364, p = 0.546), and the job type of mixed - physical (B = -0.205, SE = 0.192, Wald = 1.14, p = 0.285) were non-significant. Figure 5 plots these results.

For education, the estimated odds ratio favored a decrease of 6.2% (95% CI: 0.896, 0.982) probability of losing one point on the delayed recall questions for every year of education someone had. For job type, the estimated odds ratio favored a decrease of 35.0% (95% CI: 0.424, 0.997) probability of losing one point on the digit span questions when having an intellectual job as compared to a physical effort type job.

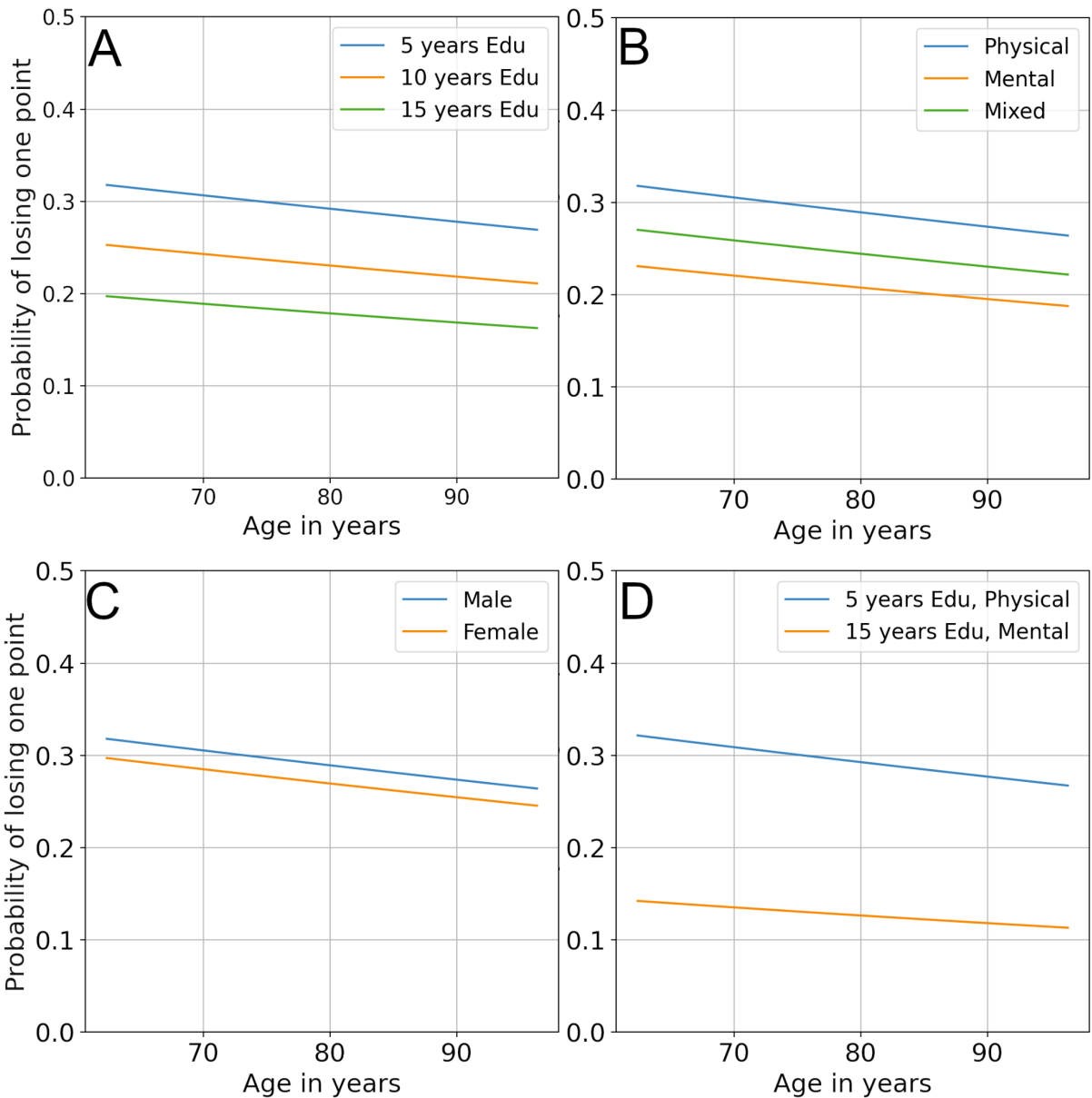

Figure 5. Probability of losing one point on the digit span working memory. Panel A shows the variability across years of education, B) across job types, C) across sex, and D) the combination of the significant effects of education and job type.

**Circle**

The model containing interaction terms was not significantly better than the model containing the main effects of the covariates ($X^2(4) = 4.18$, p = 0.382). The model with the main effects was significantly better than the model containing only age ($X^2(4) = 40.7$, $p < 0.001$) and was significant using the overall model test ($X^2(5) = 160$, $p < 0.001$). This model also had the lowest AIC of the three models.

The omnibus likelihood ratio tests showed that age ($X^2(1) = 102$, $p < 0.001$), education ($X^2(1) = 14.2$, $p < 0.001$), and job type ($X^2(2) = 10.6$, $p = 0.0051$) were significant; however, sex ($X^2(1) = 2.57$, $p = 0.109$) was not. The unstandardized estimates for the intercept ($B = -2.21$, SE = 0.164, Wald = 183, $p < 0.001$), age ($B = 0.100$, SE = 0.0104, Wald = 92.2, $p < 0.001$), education ($B = -0.134$, SE = 0.0375, Wald = 12.9, $p < 0.001$), both job types of mental - physical effort ($B = -0.848$, SE = 0.379, Wald = 5.01, $p = 0.0252$), and mixed - physical effort ($B = -0.615$, SE = 0.260, Wald = 5.59, $p = 0.0180$) were significant. The unstandardized estimates for sex ($B = 0.282$, SE = 0.177, Wald = 2.53, $p = 0.112$) was non-significant. Figure 6 plots these results.

For age, the estimated odds ratio favored an increase of 10.5% (95% CI: 1.08, 1.13) probability of losing one point on the digit span questions for every year older someone was. For education, the estimated odds ratio favored a decrease of 13.6% (95% CI: 0.812, 0.941) probability of losing one point on the delayed recall questions for every year of education someone had. For job type, the estimated odds ratio favored a decrease of 57.2% (95% CI: 0.204, 0.900) probability of losing one point on the digit span questions when having a mental effort job as compared to a physical effort type job. For job type, the estimated odds ratio favored a decrease of 46.0% (95% CI: 0.324, 0.900) probability of losing one point on the digit span questions when having a mixed effort job compared to a physical effort type job.

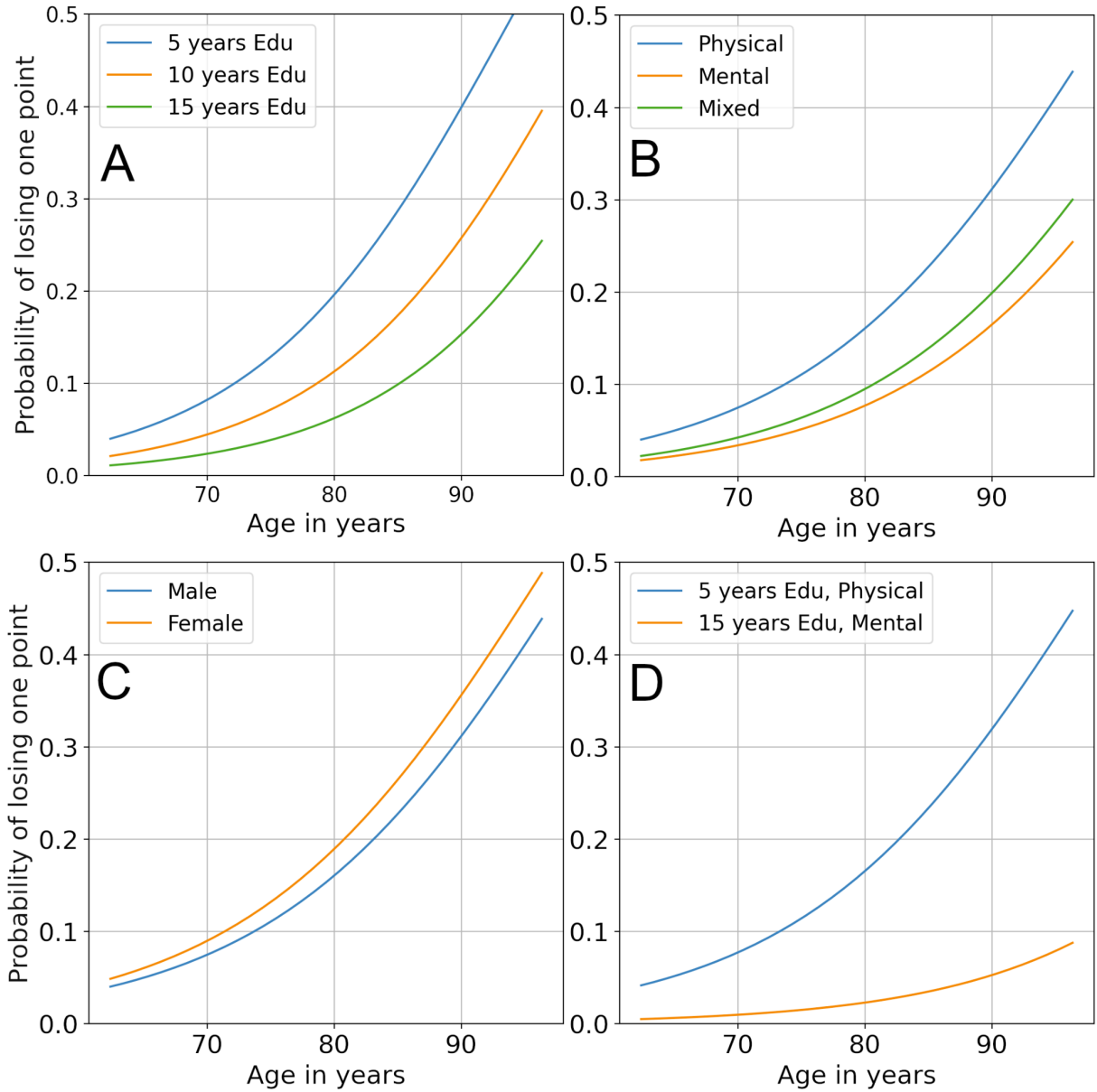

Figure 6. Probability of losing one point on the circle drawing task. Panel A shows the variability across years of education, B) across job types, C) across sex, and D) the combination of the significant effects of education and job type.

## Discussion

The results of this study demonstrate that age, education, job type, and sex all differ in their impact on cognitive performance. Advancing age consistently negatively affected task performance, which differed in its magnitude across the cognitive domains. Greater education demonstrated positive, negative, and no effects. Greater engagement in mental effort during work demonstrated consistent positive benefits compared to jobs relying more on physical effort. Sex displayed significant effects in one domain and altered the age effect in another. The results demonstrate variations in how age and lifetime exposures affect cognitive performance and the presence of additive beneficial effects among the exposures.

The effect of age demonstrated large variability across the cognitive domains. There was no significant effect of age in the immediate recall domain; in stark contrast to visuospatial abilities, assessed with the ability to draw circles, the probability of losing one point increased 50% over thirty years. The wide range of age effects demonstrated here and elsewhere (LaPlume et al., 2022) reflect the large individual variation in cognitive abilities and the likely differences in neural decline underlying cognition. However, it is also important to consider physical aging effects, such as arthritis, which were not controlled for in this study.

The effects of education also varied across the different domains. Greater years of education significantly decreased the probability of losing one point on the orientation, digit span, delayed recall, and circle tasks. In a Brazilian sample using a similar version of the MMSE, education was positively related to the orientation and circle tasks and negatively related to the delayed recall task (Laks et al., 2010). Greater education significantly negatively affected the paper task within the current sample. Those with greater education had an increased probability of losing one point. This counterintuitive finding is not novel and supports the theory of cognitive reserve (Stern, 2012). Cognitive reserve is the concept that lifetime exposures allow the brain to functionally compensate for the negative effects of aging and maintain cognitive performance. Unfortunately, this benefit comes with the downside that once cognitive decline begins, it is more rapid in those with the greatest beneficial exposures (Helzner et al., 2007). It is possible that within this sample of older adults from Barbados, the cognitive functions required to follow the instructions and perform the paper task have begun to decline before participants engaged in the current study.

Another important consideration with the use of years of education as a factor is that it does not address the type and quality of education nor historical changes to educational systems. In Barbados, the education system was structured according to the British model. School is similar for all students until age 11. At this age, students are streamed into different academic institutions based on a standardized assessment called the Common Entrance Examination. The education system in Barbados also experienced major developmental changes starting in 1967, including expanded access at primary and secondary levels and the establishment of tertiary institutions (Prepared by the Planning Research and Development Unit of the Ministry of Education, Youth Affairs and Culture, 2000). Therefore, education coded as years is a measure that ignores the effects of history, quality, type, and student stratification based on standardized test scores of academic proficiency. Future work will look at how stratification based on age 11 skills translates to the probability of a cognitive decline in late life.

High levels of mental effort required within a job were beneficial to working memory, following instructions, delayed memory recall, and visuospatial skills. Similar findings also demonstrate slower post-retirement cognitive decline (Fisher et al., 2014). Performance on the following instruction paper and delayed recall working memory tasks was significantly greater for jobs fully reliant on mental effort. In contrast, there was a beneficial mental effort dose-effect the performance of the visuospatial circle drawing task. For the backward digit span working memory task, the mental effort and mixed job types had similar benefits compared to physical effort job types. High occupational complexity related to healthy patterns of brain structure supports this result (Habeck et al., 2020; Spreng et al., 2011).

There is debate in the literature about the mechanisms underlying the benefits of occupational complexity. One idea is that high mental effort helps with physiological brain maintenance and the development of neural reserves (Cabeza et al., 2018). Another idea is that the social support and engagement required for managerial and high-mental effort employment are important for later-life cognitive functioning (Smart, 2015).

The current work uses gross labeling of job types with three categories, thereby oversimplifying the effects of occupation on one's life and aging. Future directions will use information about the occupational duration and specific occupational activities (Gadermann et al., 2014; Peterson et al., 1999). Such an approach will provide a more detailed and nuanced look at occupational demands and whether they serve as cognitive training regimens leading to maintained brain and cognitive health in later life (Habeck et al., 2020).

There was a significant sex effect for working memory where females had a greater probability of losing one point on the digit span working memory task than males. When following instructions with the paper task, sex effects show the males with an increasing probability of losing one point as they age, whereas the females do not show this increase. Taken in context with the other lifetime exposures, males have a rapid increase in the probability of losing one point as they age, which is not lessened by the beneficial effects of education and job type. On the other hand, the females do not demonstrate a rapid increase in the probability of losing one point as they age, even for the lower educated, physical effort-based workers.

The same "following instructions" paper task was also a strong predictor in a prospective cohort study of depression and dementia (Berger et al., 2005). Within dementia prevalence, females are more likely to develop dementia than males ("2022 Alzheimer's Disease Facts and Figures," 2022). Therefore, the current work demonstrates that one of the strongest individual predictors of dementia when using the MMSE is modulated by lifetime exposures in females. This finding provides hope that the promotion of greater education and mental effort occupations for females will benefit their well-being and help delay the onset of cognitive decline. This finding also highlights the differential effectiveness of cognitive reserve mechanisms between the sexes.

The current analyses employed sex, education, and job type in the same model, allowing the examination of individual and additive effects between factors. Additive effects are evident in all cognitive domains having significant findings. Therefore, the lifetime exposures utilized in this work demonstrate interrelations with each other where their combinations provide greater cognitive benefits than anyone alone. Future work will investigate the additive effects to explore the optimal recipe for healthy aging.

Interestingly, the finding of additive lifetime exposure effects addresses a concern in the cognitive reserve literature. Cognitive reserve is the concept that lifetime exposures decrease the negative effect of age and disease-related pathology on cognitive performance. An open question is how combinations of lifetime exposures should be modeled and studied. Should their common effects of lifetime exposures be reflected in a global measure of cognitive reserve (Nucci et al., 2012), or do individual contributions of various lifetime exposures combine to form cognitive reserves (Jones et al., 2011)? The current findings of additive effects between education, job type, and sex support the latter idea.

The mechanisms underlying the beneficial effects of lifetime exposures include brain and cognitive reserve concepts. The idea is that the experiences one encounters due to sex, education, and employment provide individuals with the neural and cognitive support structures to minimize the negative effects of aging and neural disease on cognitive performance. This minimization may be physiological such that certain exposures decrease the negative effect of age on brain tissue or that brain tissue is better maintained (Cabeza et al., 2018; Steffener et al., 2014). The effects of lifetime exposures may also provide the means for someone to develop cognitive skills that transfer to better test performance. The current work focused on identifying factors that moderate the age effects themselves. Only sex with the following instructions paper task altered the age effects in the current sample. The rest of the factors used in this study directly altered cognitive performance. The result is that clinically relevant age-related declines in cognitive performance may be delayed.

Within Barbados, high levels of education are an important source of pride for the nation. Among the seven nations that participated in the larger SABE study, Barbados and Cuba had significantly higher years of education than the other nations (Maurer, 2011). The current work demonstrates the value of this focus. Higher education decreased the probability of cognitive decline with advancing age. Higher education allows for greater job opportunities where the jobs exist. Unfortunately, on an island such as Barbados, job opportunities may still be limited to those relying on physical effort, even for individuals with advanced degrees. Despite any such limitations, the current results demonstrate that when higher education, coupled with employment requiring high levels of mental effort, the two factors beneficially combine to lessen age-related cognitive decline.

An additional factor in potential beneficial lifetime exposures in Barbados is the widespread use of both a Bajan dialect and the use of English. Therefore, the conduction of daily life involves repeated switching between the two and is considered a form of bilingualism (Oschwald et al., 2018), with active engagement having beneficial cognitive effects (Bialystok, 2015). Although not addressed due to a lack of questions about language use in the study, future work will explore whether using a dialect provides cognitive benefits. Greater understanding in this domain provides public policy insight into preserving local dialects' cultural and cognitive benefits.

One of the largest limitations of this work was using the nineteen-point MMSE as a cognitive assessment tool. The MMSE was designed as a screening tool for dementia, and although it assesses multiple cognitive domains, it does so in a limited fashion. Future directions would involve more comprehensive cognitive assessments with tools that do not have strong ceiling effects as with the MMSE. Expanded cognitive assessments allow more nuanced and detailed assessments of cognitive differences across ages, occupations, education, and sex.

Despite a relatively large sample size of over 1300 participants, this was insufficient to address many interesting questions. Only three moderating variables were included in the modeling: sex, job type, and years of education. The dataset also includes measures of engagement in exercise, social, and artistic leisure activities. An ideal analysis would include a wide array of lifetime exposures to identify patterns of lifestyles that have the greatest benefit for minimizing the effects of age on cognition. However, much larger sample sizes are needed to ensure adequate sample sizes in all cells of such a model. For instance, 45 percent of the current sample stated they exercised regularly. Within the context of delayed recall, the factor with the most number of people demonstrating a decline, there are eight male participants with intellectual jobs that exercise regularly. For immediate recall, there are no participants in that cell. For these reasons, the current analyses utilized a limited number of lifetime exposures.

The current work did not test for sex, education, and job type interactions. There is evidence that being female limits one's access to educational opportunities, leading to lower-status job types (Mandel & Semyonov, 2006). This sex effect predicts a higher probability of losing points on cognitive assessments for females than males as people age. Unfortunately, the sample size and variability in the MMSE cognitive assessment limit the ability to explore such effects.

## Conclusion

Currently, there is a lack of information about cognitive decline in the Caribbean. This study demonstrated that significant findings were found even within the limited scope of the MMSE cognitive screening tool. Age is not a universal negative process affecting all aspects of cognitive performance. In addition, adverse age effects may be minimized or delayed through fixed and modifiable lifetime exposures. Of great interest is that the benefits of lifetime exposures are additive, providing combined benefits. Follow-up studies with detailed assessments of cognitive function and beneficial lifetime exposures will provide insight into the prevalence of cognitive decline in Barbados. They will identify the beneficial lifetime exposures that people engage in, providing insight into public policy planning for handling increases in the number of adults with cognitive decline and dementia.